\documentclass[twocolumn,prl,amsmath,amssymb,showpacs,superscriptaddress,floatfix]{revtex4}
\usepackage{graphicx}
\usepackage{bm}
\usepackage{graphicx}

\newcommand{\refe}[1]{(\ref{#1})}
\newcommand{\bvec}[1]{\mathbf{#1}}
\renewcommand{\vec}[1]{\bvec{#1}}
\renewcommand{\phi}{\varphi}

\newcommand{\bra}[1]{\langle#1\!\mid}
\newcommand{\ket}[1]{\mid\!#1\rangle}

\newcommand{\zero}{\phi_{0,0}}
\newcommand{\one}{\phi_{1,0}}

\begin{document}

\title{Non-local state-swapping of polar molecules in bilayers}

\author{A. Pikovski$^{1}$, M. Klawunn$^{2}$, A. Recati$^{2}$, and L. Santos$^{}$}
\affiliation{\mbox{$^{}$Institut f\"ur Theoretische Physik, Leibniz Universit\"at Hannover, Appelstr. 2, 30169, Hannover, Germany}\\
\mbox{$^{2}$INO-CNR BEC Center and Dipartimento di Fisica, Universit\`a di Trento, 38123 Povo, Italy}
}
\date{\today}

\begin{abstract}
The observation of significant dipolar effects in gases of ultra-cold polar molecules typically demands a strong external electric field 
to polarize the molecules. We show that even in the absence of a significant polarization, dipolar effects may play a crucial 
role in the physics of polar molecules in bilayers, provided that the molecules in each layer are initially prepared in a different 
rotational state. Then, inter-layer dipolar interactions result in a non-local swap of the rotational state between molecules 
in different layers, even for weak applied electric fields.
The inter-layer scattering due to the dipole-dipole interaction leads to a non-trivial dependence of the swapping rate 
on density, temperature, inter-layer spacing, and population imbalance. 
For reactive molecules like KRb, chemical recombination immediately follows a non-local swap and dominates the losses 
even for temperatures well above quantum degeneracy, and could be hence observed under current experimental conditions.
\end{abstract}


\maketitle


A new generation of experiments has started to explore the remarkable novel physics
of dipolar gases, in which dipole-dipole interactions play a key role~\cite{Lahaye2009}. 
These interactions, being long-range and anisotropic, differ from
the short-range isotropic interactions which have dominated up to now
ultra-cold atomic physics. Polar molecules are expected to provide fascinating 
new scenarios for quantum gases due to their large electric dipole moments.  
Recent experiments on preparation and control of ro-vibrational 
and hyperfine states of KRb at JILA~\cite{Ni2008} open new 
perspectives towards a degenerate quantum gas of polar molecules. Unfortunately
chemical recombination due to the reactive character of KRb has 
up to now prevented to reach quantum degeneracy~\cite{Ospelkaus2010}.
However, dipolar interactions between partially polarized KRb molecules have been recently shown to  
significantly reduce chemical recombination in constrained  
quasi-2D geometries~\cite{DeMiranda2011}. 

Although a polar molecule in its lowest ro-vibrational state may  
have a large dipole moment in the molecular frame, in the absence of 
an external electric field the dipole moment in the laboratory frame averages to zero. 
As a consequence, the observation of dipolar effects in gases of polar molecules
typically demands the polarization of the molecules in an external electric field~\cite{Lahaye2009}.
For example, a KRb molecule in its singlet ground state has a permanent dipole moment of
$d=0.566$ Debye in the molecular frame~\cite{Ni2008}, but rather large fields of several kV$/$cm must be employed 
to reach effective dipoles of $\sim 0.2$ Debye in the lab frame~\cite{Ni2010}. 

However, remarkably, dipolar interactions may play a significant role even in the absence of an external electric field, 
if the molecules are prepared in different rotational states~\cite{Barnett2006}. In that case, the interaction 
between molecules in different rotational states is produced by exchanging a quantum of angular momentum~(state swap), 
resembling the resonant interaction of an electronically excited atom and a ground state one.

\begin{figure}
\vspace{-0.7cm}
\begin{center}
\includegraphics[width=0.3\textwidth,angle=0]{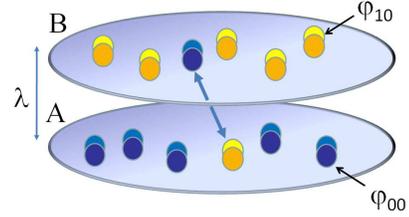}
\vspace{-0.6cm}
\caption{Bilayer set-up with molecules in rotational state $\zero$~(layer A) and $\one$~(layer B).
A state swap has been sketched.}\label{fig:1}
\end{center}
\vspace*{-0.8cm}
\end{figure}

In this Letter, we show that bilayer systems may provide an excellent 
environment for observing the non-trivial physics associated with dipole-induced state-swaps in current experiments.
We consider polar molecules in a bilayer geometry, without inter-layer tunneling, prepared such that  
the molecules in layer $A$~($B$) are in the  
ground~(first-excited) rotational state~(Fig.~\ref{fig:1}). 
The dipole-dipole interaction induces  
a molecule in one layer to interchange its rotational state with a molecule in a different state in  
the other layer, as sketched in Fig.~\ref{fig:1}. Such a process resembles 
spin-exchange in the collision of polarized atomic beams~\cite{Glassgold1963} and  
spin-changing collisions in spinor gases~\cite{Ueda2010}. 
We show that the special features of the inter-layer scattering lead, especially for quantum degenerate samples, 
to a non-trivial dependence of the swap rate on temperature,  
molecular density, inter-layer spacing, and population imbalance.  Interestingly, 
for reactive molecules like KRb chemical recombinations following a state-swap 
dominate losses even well inside the non-degenerate regime and could be observable 
under current experimental conditions.

We are interested in fermionic polar molecules confined in two thin
parallel layers, $A$ and $B$, separated by a distance $\lambda$ along $z$, and with negligible
hopping between them~(Fig.~\ref{fig:1}). 
We consider diatomic molecules with a permanent electric dipole moment $d$
in the electronic singlet state ($^1\Sigma$) and in the vibrational ground state. 
A weak external electric field ${\bf E}={\mathcal E}\hat{\bf  z}$ is applied.
The rotational states of the molecules are described by the Hamiltonian
\begin{equation}
H_\text{mol}=B{\bf J}^2-{\bf d}\cdot{\bf E},
\label{Hmol}
\end{equation}
where $B$ is the rotational
constant, ${\bf  J}$ is the angular momentum
operator, and ${\bf d}=d(\sin\theta\cos\varphi,\sin\theta\sin\varphi,\cos\theta )$ is the
dipole moment operator, where the angles $\theta$ and $\varphi$ describe the orientation of the molecule 
in the laboratory frame.
For a weak electric field, such that $\beta=d{\mathcal E}/B\ll 1$, the
eigenstates $\varphi_{J,M}$ of the Hamiltonian~(\ref{Hmol}) are still well described by 
the quantum numbers $J$ and $M$ associated with the angular momentum ${\bf J}$, and they can be found 
using perturbation theory~\cite{Herzberg1953,Micheli2007}. 
The electric field lifts the degeneracy 
of the three $J=1$ states, which become separated by an energy 
$\Delta E=E_{1,0}-E_{1,\pm 1}=3 \beta^2 B/20$.
The effective dipole moment in the laboratory frame, $\langle {\bf d} \rangle$, is very small for a weak
field. For example, the effective dipole in the rotational ground state is
$\langle \varphi_{0,0}|{\bf d}|\varphi_{0,0}\rangle = \beta d /3 \,\hat{\bf z}$, 
giving $0.019$ Debye for KRb and $\beta=0.1$, corresponding to ${\mathcal E} = 0.4\,{\rm kV}/{\rm cm}$;
this results in a dipolar interaction five times smaller than that for atomic magnetic dipoles in Chromium~\cite{Lahaye2009}. 

We suppose that, initially, the molecules in layer $A$ are prepared in the state $\zero$, while 
those in $B$ are prepared in $\one$, as illustrated in Fig.~\ref{fig:1}. 
Such selective preparation may be achieved by applying a spatially
varying electric field, and then transferring the molecules
of layer $B$ into the desired state with a microwave transition~\cite{Jin-Ospelkaus}. 
The dipole-dipole interaction between two molecules with dipole 
${\bf d}_{1,2}$ at positions ${\bf r}_{1,2}$ is
\begin{equation}
H_{DD}= \frac{1}{4\pi\epsilon_0} \left[ \frac{\bvec{d}_1 \cdot \bvec{d}_2}{|\vec{r}|^3} - \frac{3(\vec{d}_1 \cdot \vec{r})(\vec{d}_2 \cdot \vec{r} )}{|\vec{r}|^5} \right],
\end{equation}
with ${\bf r}={\bf r}_1-{\bf r}_2$.
Since the anisotropic dipole-dipole interaction does not preserve the rotational quantum number, a collision may lead to 
an interchange of the rotational state of molecules in different layers:
\begin{equation}
\zero^A \: \one^B \to \one^A \: \zero^B .
\label{stateswap}
\end{equation}
The matrix elements
of $H_{DD}$ involving the states $\zero$ and $\one$ are given by, neglecting terms $O(\beta^2)$,
\begin{align}\label{eq:DDI_0}
\!\bra{\one^A \, \zero^B} \! H_{DD} \! \ket{\one^A \, \zero^B} \! &=0,\\
\!\bra{\one^A \, \zero^B} \! H_{DD} \! \ket{\zero^A \, \one^B} \! &=\!
\frac{d^2}{3r^3} \! \left[ 1 -  \frac{3 z^2}{r^2} \right] \equiv V_{dd}({\bf r}).\!\!
\label{eq:DDI}
\end{align}
Transitions into rotational states other than $\zero$ and $\one$ can 
be neglected. They involve an energy change of $\Delta E$, which is much larger
than other energy scales in the system (the rotational constant $B$ is typically very large), 
such as the dipolar energy $E_D=d^2/(4\pi \epsilon_0 \lambda^3)$.
For KRb, for example, $B/h = 1.11$ GHz~\cite{Ni2008}, thus $\Delta E / k_B= 80 \mu{\rm K}$ is
much larger than $E_D/k_B = 15 n{\rm K}$
for $\lambda=532 {\rm nm}$.
Therefore, we only consider processes between the states $\zero$ and $\one$
\footnote{%
A sufficiently large electric-field gradient $d{\mathcal E}/dz$ may inhibit state swapping.
This may be neglected if $E_D \gg \frac{2\beta d\lambda}{15} \frac{d{\mathcal E}}{dz}$, 
giving $d{\mathcal E}/dz\ll 0.5 $kV$/$cm$^2$ for $E=0.4$kV$/$cm.}.

At this point the dipolar interaction is easily diagonalized using 
the symmetric and antisymmetric two-body states
$|S,A\rangle \equiv ( |\zero^A \, \one^B \rangle \pm  |\one^A \, \zero^B \rangle ) / \sqrt{2}$, 
for which
$\langle A |H_{DD}|A\rangle = - V_{dd}({\bf r})$, 
$\langle S |H_{DD}|S\rangle = V_{dd}({\bf r})$, and 
$\langle S |H_{DD}|A\rangle = 0$.
The inter-layer scattering is hence decomposed into two separate channels, namely  
the symmetric one, with interaction potential $V_{dd}({\bf r})$, and the antisymmetric one, with $-V_{dd}({\bf r})$. 
It may be noted that in analogy with spin-exchange collisions between polarized atomic beams~\cite{Glassgold1963},
the interaction between two molecules 
can be written as $V=V_{dd}({\bf r}) P_S - V_{dd}({\bf r}) P_A $, where
$P_S, P_A$ are the projection operators for states $|S\rangle, |A\rangle$.

We introduce the 
scattering amplitudes $f_+$ and $f_-$, which characterize the low-energy inter-layer scattering for the symmetric and antisymmetric channel, respectively. 
The scattering amplitude for the state-changing process in Eq.~(\ref{stateswap}) is then given by
\begin{equation}
f_\text{sc}(k) = \tfrac{1}{2} (f_+(k) - f_-(k)),
\end{equation}
with $\hbar k$ the relative momentum between the colliding particles.
The amplitude for state-preserving collisions, $\zero^A \: \one^B \to \zero^A \: \one^B$,  
is $f_\text{sp}=\frac{1}{2}(f_- + f_+)$.

\begin{figure}[t]
\vspace{-0.2cm}
\begin{center}
\raisebox{-3ex}{(a)}\includegraphics[width=0.26\textwidth,angle=270]{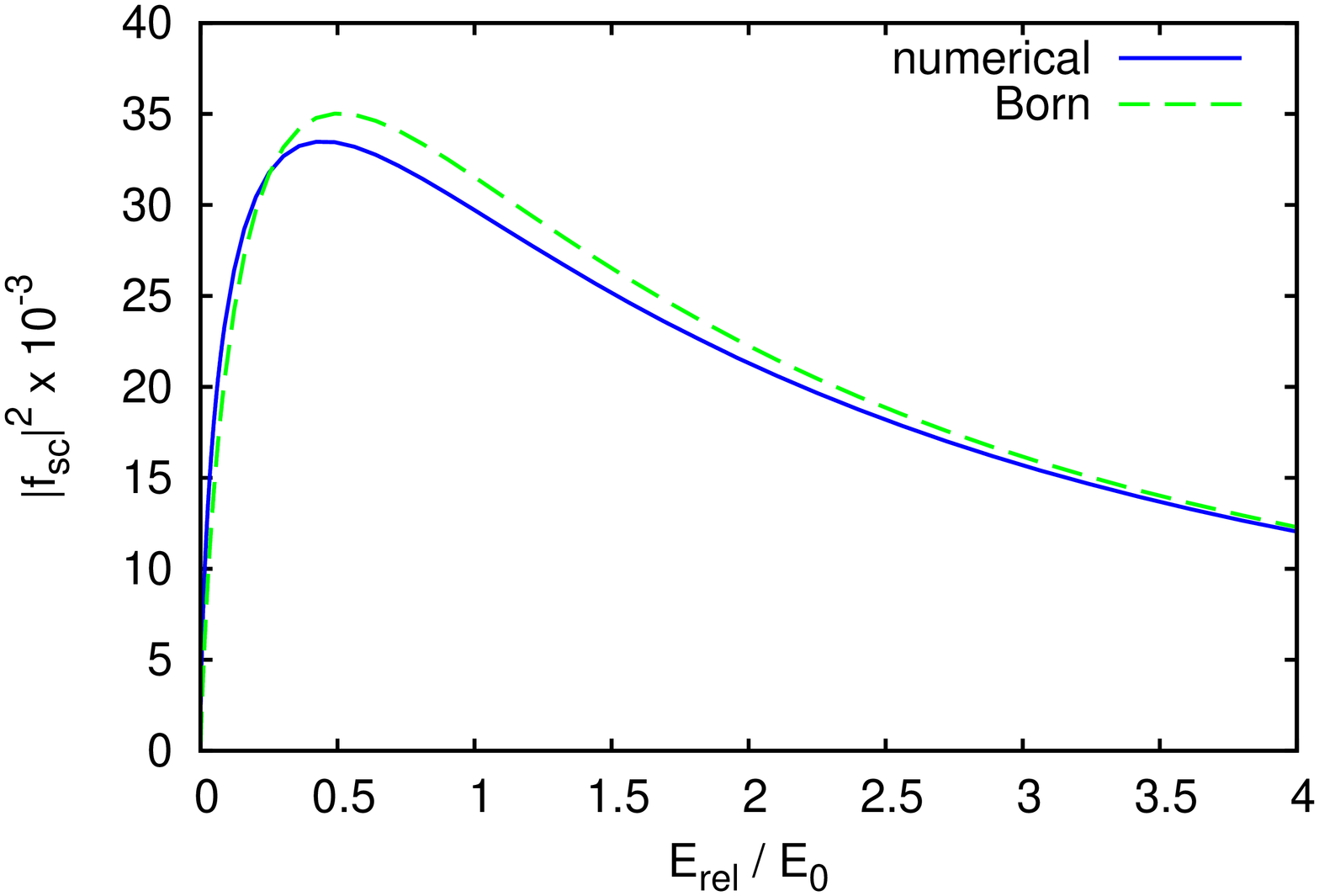}
\raisebox{-3ex}{(b)}\includegraphics[width=0.27\textwidth,angle=270]{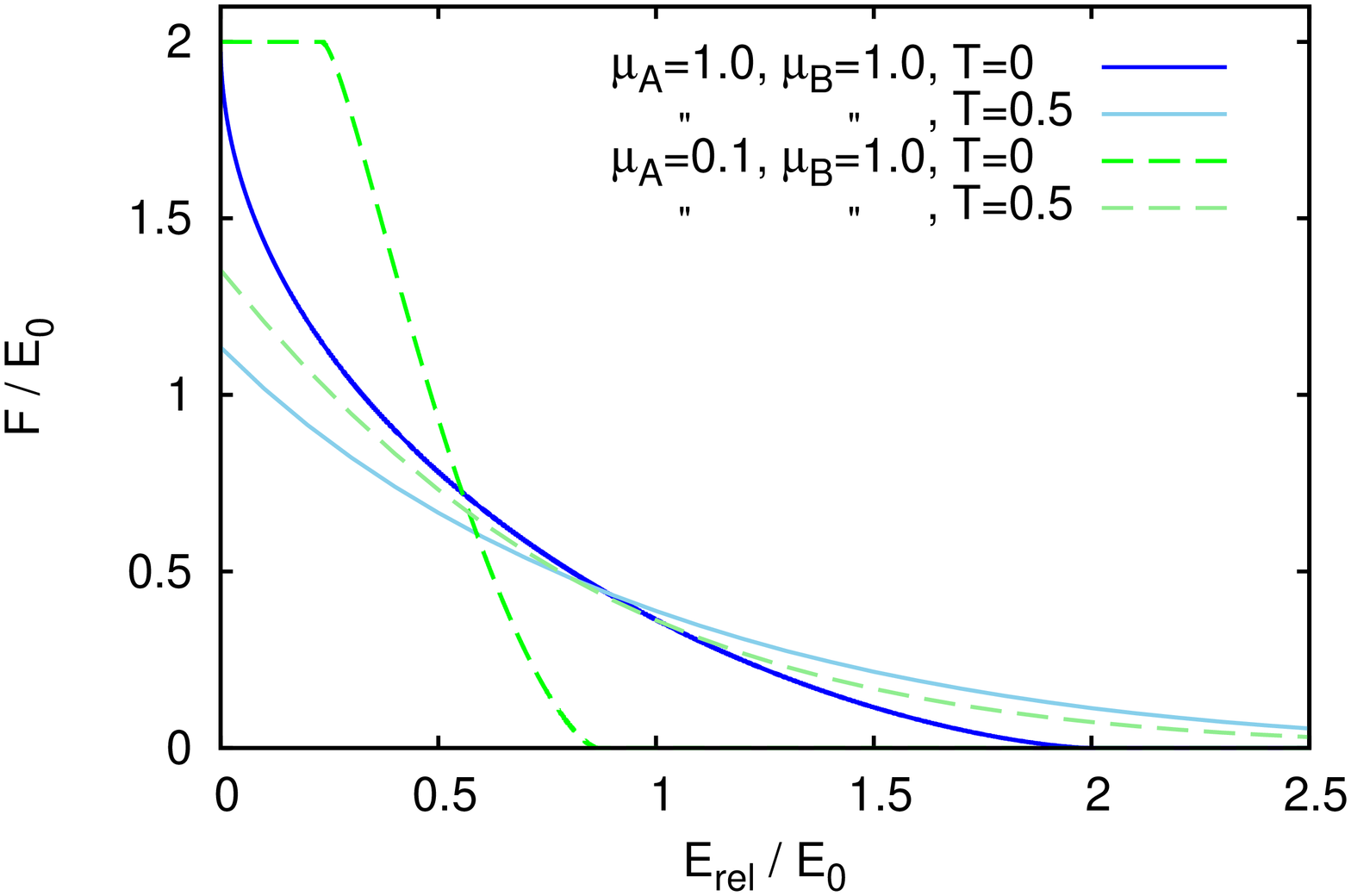}\vspace*{-1em}
\caption{a) Scattering amplitude for state-changing collisions. 
b) Distribution of relative kinetic energies $F(E_\text{rel})$ for different values of $\mu_A$, $\mu_B$, and $T$ (in units of $E_0$).}
\label{fig:2}
\end{center}
\vspace*{-0.8cm}
\end{figure}

Collisions involving particles in different layers can be described here by two-dimensional
scattering ($s$-wave only).
The amplitudes $f_+$ and $f_-$ 
are found by numerically solving the 2D Schr\"odinger equation
with the potential 
$\pm V_{dd}(x,y,z\!=\!\lambda)$.  
This may be compared, using Ref.~\cite{Klawunn2010}, 
to the result from the Born approximation for $f_\pm$,
\begin{equation}\label{eq:f-born}
|f_\text{sc}(q)|^2 = U_0^2 q^2 \left [ 
2+ \pi({\bf L}_1(2q)-I_1(2q))
\right ]^2 + {\cal O}(U_0^4),
\end{equation}
with $q=k\lambda$ and ${\bf L}_1(x)$ ($I_1(x)$) the modified Struve (Bessel) function.
We use the unit of energy $E_0=\hbar^2/(m \lambda^2)$, with $m$ being the mass of the molecule,
and the dimensionless coupling strength 
$U_0= E_D / (3 E_0)$.
To be specific, we will take the values for KRb and $\lambda = 532{\rm nm}$ for the plots, 
so $E_0/k_B = 13.5 {\rm nK}$ and $U_0=0.38$.
In Fig.~\ref{fig:2}(a) we show that the expression in Eq.~\eqref{eq:f-born} is in good agreement 
with the numerical solution of the Schr\"odinger equation, for the regime of interest.

Note that $|f_\text{sc}|^2$ presents a non-monotonical dependence 
with the relative kinetic energy, and falls off slowly for large energies, see Fig.~\ref{fig:2}(a).
Interestingly, the swapping amplitude in Eq.~(\ref{eq:f-born}) is 
$1/3$ of the scattering amplitude expected for fully polarized molecules
calculated in the same approximation. 
On the other hand, the scattering amplitude for state-preserving collisions is much smaller.
Since there is no direct process (the matrix element vanishes, Eq.~(\ref{eq:DDI_0})), these 
collisions are of higher order: $|f_\text{sp}|^2={\cal O}(U_0^4)$.

State-swapping leads to molecules in $\zero$~($\one$) in layer $B$~($A$), see Fig.~\ref{fig:1}.
For short times, the number density $n_{d}$ of these ``defects'' satisfies the rate equation
\begin{equation}
\label{eq:Rate}
d n_d/dt= K n_A n_B,
\end{equation}
where $n_A$ and $n_B$ are the densities in layers $A$ and $B$. 
The rate $K$ is determined by the cross-section for state-changing collisions
$\sigma_{sc}$, averaged over the velocity distributions $\bar{f}_{A,B}$ of particles in 
layer $A$ and $B$~\cite{Shuler1968}: 
$K = \left\langle v \sigma_\text{sc} \right\rangle 
 = \iint v \sigma(p) \bar{f}_A(\vec{v}_A) \bar{f}_B(\vec{v}_B) \, d^2v_A d^2v_B$,
where $\vec{v}=\vec{v}_A-\vec{v}_B$ is the relative velocity for a pair of molecules
and $\vec{p}=(m/2)\vec{v}$. The relation to the scattering amplitude in two dimensions is
$v \sigma = 4 \frac{\hbar}{m/2} |f|^2$.

For short times or for a small defect density, Pauli-blocking effects can be neglected.
For reactive molecules, in fact, the defect density is always small because
swapped molecules are quickly lost, as discussed below.
For equilibrium distributions which depend only on energy, the collision rate $K$ becomes:
\begin{equation}
K = 8\frac{\hbar}{m}\int_0^\infty  |f_\text{sc}(k_\text{rel})|^2 \, F(E_{\text{rel}}) d E_{\text{rel}},
\label{eq:K}
\end{equation}
where $E_\text{rel}=\hbar^2k_\text{rel}^2/m=\frac{1}{2} (m/2) v^2$.
Here
\begin{equation}
F(E_{\text{rel}}) =  \int_0^{2\pi} \frac{d\gamma}{2\pi} \int_0^\infty f_A (E_A) f_B(E_B) dE_{\text{cm}} 
\label{F}
\end{equation}
is the distribution of relative kinetic energies, where
$\cos \gamma=\frac{\bvec{v} \cdot \bvec{V}}{v V}$,
$\bvec{V}=(\bvec{v}_A + \bvec{v}_B)/2$, 
$E_{A,B}=mv_{A,B}^2/2$, and 
$E_{\text{cm}}=m V^2$ is the center-of-mass kinetic energy.
The $f_{A,B}$ are normalized to $\int_0^\infty f(E) dE =1$.

The distribution of relative energies $F(E)$ can be calculated from Eq.~\refe{F}, assuming
a Fermi distribution in each layer $f_{A,B}(E)={\mathcal N}^{-1} (e^{(E-\mu(T))/k_BT}+1)^{-1}$,
with the normalization constant ${\mathcal N}=T \log(1+e^{\mu/k_BT})$.
The chemical potential $\mu(T)$ is that of the ideal 2D Fermi gas, i.e.\ the solution of
$\log (1+e^{\mu/k_BT}) = \epsilon_F / k_BT$
for a given temperature and a given Fermi energy $\epsilon_F=\mu(0)=2\pi n \hbar^2/m$,
where $n$ is the molecular density in the layer.
Figure~\ref{fig:2}b shows $F(E)$ for different temperatures.
For temperatures $ k_B T \gg \epsilon_F$, the relative energies follow a Boltzmann distribution, $F(E)=1/k_B T \exp (-E/k_B T)$.

\begin{figure}[t]
\vspace*{-0.2cm}
\begin{center}
\raisebox{-3ex}{(a)}\includegraphics[width=0.27\textwidth,angle=270]{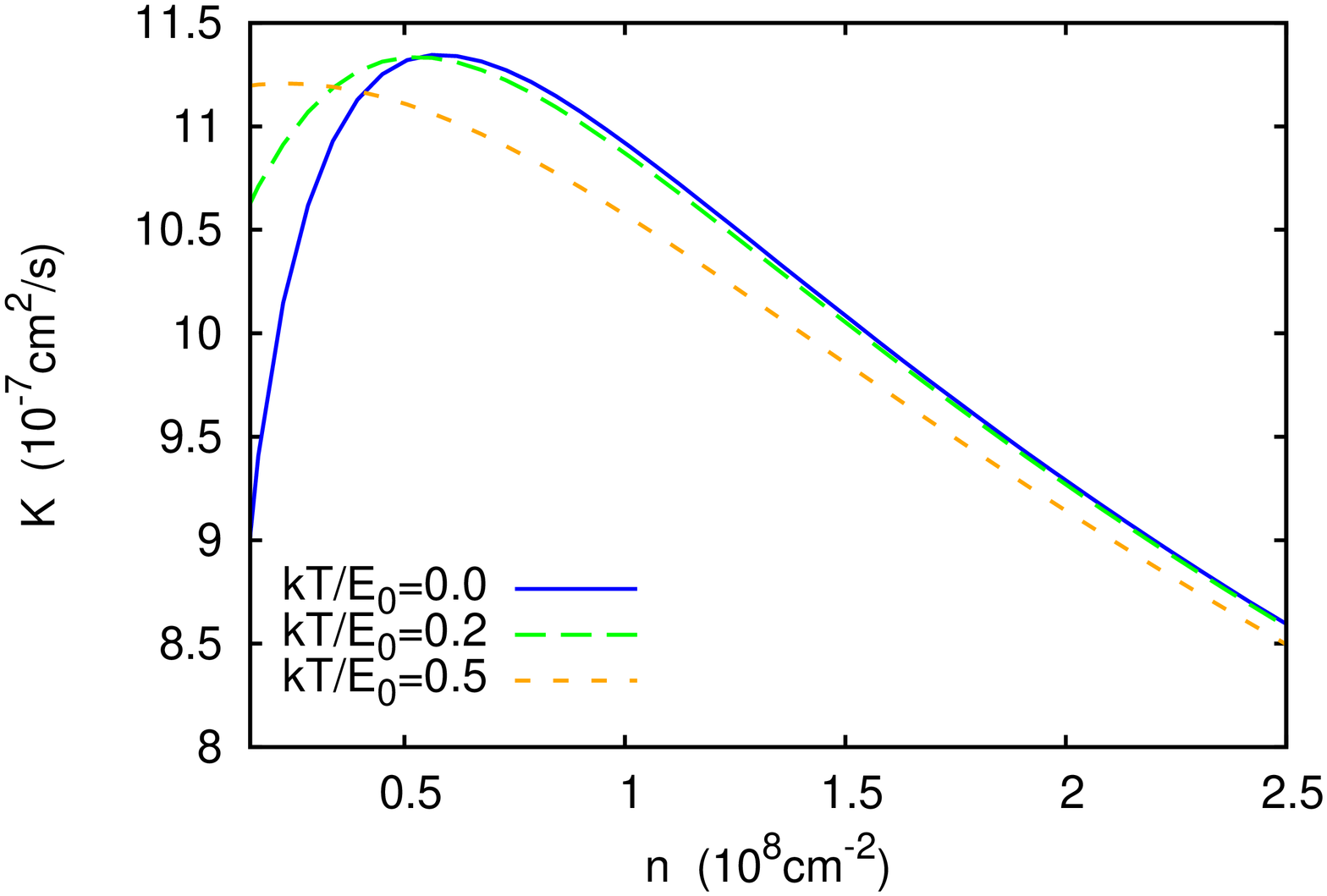}
\raisebox{-3ex}{(b)}\includegraphics[width=0.27\textwidth,angle=270]{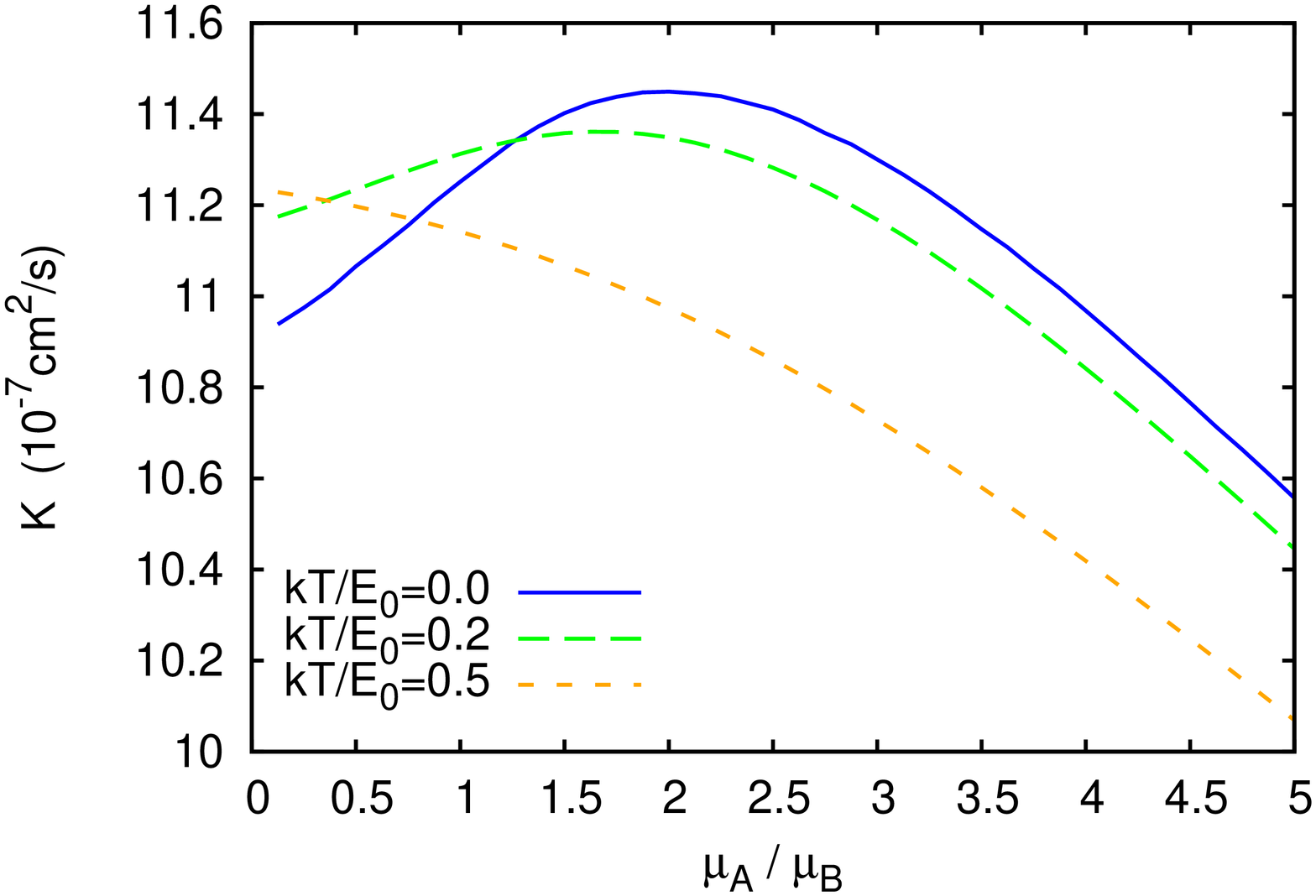}\vspace*{-1em}
\caption{Swapping rate $K$ for different $T$ as a function of: (a) the density $n$
and~(b) the imbalance $\mu_A/\mu_B$ for $\mu_B=0.8 E_0$. }
\label{fig:3}
\end{center}
\vspace*{-0.8cm}
\end{figure}

The swap rate $K$, as given by Eq.~\eqref{eq:K}, is the overlap integral
between $F(E_\text{rel})$ and $|f_\text{sc}(E_\text{rel})|^2$. 
The dependence of the collision rate on the density, inter-layer spacing $\lambda$, and temperature
can be obtained by analyzing this overlap.
The scattering amplitude $|f_\text{sc}|^2$ has a maximum around 
$E_\text{rel}/E_0=(\lambda k_\text{rel})^2 = 0.5$, see Fig.~\ref{fig:2}a, so
at $T=0$ the maximal swapping rate occurs for Fermi momenta $k_{F}\lambda\sim 1$. 
Hence, Fermi degenerate molecular samples show a 
peculiar non-monotonical behavior of the swapping rate as a function of 
density, or inter-layer spacing, as shown in Fig.~\ref{fig:3}(a). 
This effect is visible if $k_B T \ll E_0$, and it is smeared for higher temperatures, see Fig.~\ref{fig:3}(a).

The swap rate $K$ presents as well an interesting dependence on the population imbalance between both layers.
At $T=0$ the function $F(E)$, which may be analytically evaluated, depends on the chemical potentials 
$\mu_A$ and $\mu_B$ of the layers (or, equivalently, particle densities), 
acquires the form of a smoothed step function, see Fig.~\ref{fig:2}(b).
Interestingly, as shown in Fig.~\ref{fig:3}(b), the swap rate $K$ may show a maximum as a function of the imbalance 
$\mu_A / \mu_B$, if $\mu_B \lesssim E_0$. Again, as can be seen in Fig.~\ref{fig:3}(b), at finite $T$, the maximum persists 
although smeared out.

\begin{figure}[!t]
\vspace{-0.2cm}
\begin{center}
\includegraphics[width=0.27\textwidth,angle=270]{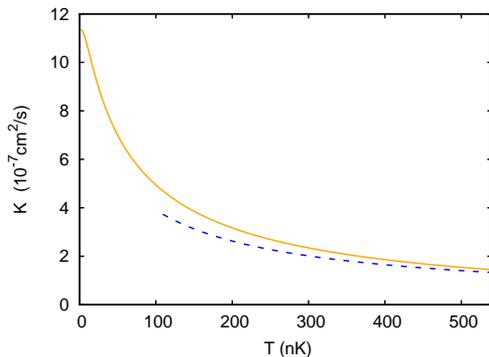}\vspace*{-1em}
\caption{Swapping rate $K$ as a function of temperature, for $\epsilon_{F}/E_0=1$.
The dashed line is the result of Eq.~(\ref{largeT}).
}
\label{fig:4}
\end{center}
\vspace*{-0.8cm}
\end{figure}

The previous results suggest that the state swapping collisions may lead to very relevant effects even in the high-temperature regime,
especially in the context of reactive molecules. 
Typical experimental values for $E_0/k_B$ and $\epsilon_F/k_B$ are in the range of 
$10$nK, while
currently available experimental temperatures are much higher, namely in the range of several $100$nK. 
In the following we discuss 
the particularly relevant scenario of KRb molecules, assuming a 
molecular density of $5.6\times 10^7$cm$^{-2}$, which gives $\epsilon_F/E_0=1.0$.
Figure~\ref{fig:4} shows the temperature dependence of the swapping rate under these conditions. 
For $k_B T\gg \epsilon_F$, it follows from Eq.~(\ref{eq:f-born}) that the swapping rate acquires the form 
\begin{equation}\label{largeT}
K(T) \simeq 2\frac{\hbar}{m} U_0^2 \frac{\log (k_B T/E_0)}{k_B T/E_0}.
\end{equation}

Once chemically reactive molecules collide within one layer, they undergo a reaction of the form $\text{KRb} \to \text{K}_2 + \text{Rb}_2$ 
and leave the trap~\cite{DeMiranda2011}.
For weak electric fields (no significant polarization of the dipoles),
this results in a loss rate of
$K_{p}\simeq 2\times 10^{-7}$cm$^2/$s for $T=800$nK due to $p$-wave scattering, scaling linearly with $T$ due to Wigner threshold law~\cite{DeMiranda2011}. 
The swapping process discussed here leads to an additional loss mechanism for reactive molecules, 
since the creation of defect molecules (state $\zero$ in layer $B$ and state $\one$ in layer $A$, see Fig.~\ref{fig:1}) 
allows for intra-layer $s$-wave collisions. Their typical rates are
$K_{s}\simeq 10^{-5}{\rm cm}^2/{\rm s}$~\cite{DeMiranda2011}, i.e. at least one order of magnitude larger than the 
swapping rate, see Fig.~\ref{fig:3}.
Since $K_{s}\gg K$, the defect molecules are quickly lost (the instantaneous 
number of defect molecules at short times remains at any time very small $n_{d}\simeq (K/K_{s})n_A$). 
Hence we may equate the swapping rate with molecular loss rate, 
$dn_A/dt \simeq -(K+K_{p})n_A^2$, assuming balanced mixtures with $n_A=n_B$. 
Note that the swapping mechanism discussed here becomes the dominant loss mechanism at low-enough $T$ 
characterized by $K_{p}\ll K$. Since
swap-induced losses decrease with $T$ while losses associated to $p$-wave scattering increase linearly with $T$,
there is a temperature threshold $T_c$ at which both loss mechanisms are equally relevant. 
From Fig.~\ref{fig:3} and Ref.~\cite{DeMiranda2011}, we estimate this threshold at approximately $T_c \simeq 530$nK$\simeq 40 \epsilon_F/k_B$ 
\footnote{%
A Zeno-like effect, similar to that discussed by N. Syassen {\it et al.}, Science {\bf 320}, 1329 (2008),
may reduce the swapping rate.
We expect this correction to be small due to the moderate $K_s/K$ ratio.}.
Therefore even in the absence of external polarization, non-local dipole-induced swapping dominate 
over $p$-wave losses even deeply inside the non-degenerate regime.

In summary, polar molecules in bilayer systems in a different rotational state 
in each layer provide a distinctive setting for the study of non-local state-swaps. 
We have shown that the special features of the inter-layer dipole-dipole 
interaction lead to a non-trivial dependence of the swap rate with the parameters of the system, especially 
within the degenerate regime. Moreover, for chemically reactive molecules, the swapping leads to losses, since a
swap is followed by a quick recombination due to an intra-layer $s$-wave collision.
As a result of that, non-local state-swapping should lead to significantly 
enhanced losses below a temperature threshold, which lies well inside the non-degenerate regime.
Hence, our results show that strong non-local dipolar effects should be observable in current experiments 
in the absence of a significant external polarization.

We thank 
G. Ferrari, D. Jin, S. Ospelkaus, G.~V. Shlyapnikov, and Y. Ye 
for enlightening discussions. 
We acknowledge support from the Center for Quantum Engineering and Space-Time
Research QUEST, the DFG (SA 1031/6), and the ERC (QGBE grant).



\begin{thebibliography}{99}


\bibitem{Lahaye2009}  T. Lahaye {\it et al.}, Rep. Prog. Phys. {\bf 72}, 126401 (2009).

\bibitem{Ni2008} K.-K. Ni {\it et al.},
Science {\bf 322}, 231 (2008).

\bibitem{Ospelkaus2010} S. Ospelkaus {\it et al.},
Science {\bf 327}, 853 (2010).

\bibitem{DeMiranda2011} M. H. G. de Miranda {\it et al},
Nature Physics {\bf 7}, 502 (2011).

%
%

\bibitem{Ni2010} K.-K. Ni {\it et al.},
Nature {\bf 464}, 1324 (2010).

\bibitem{Barnett2006} R. Barnett, D. Petrov, M. Lukin and E. Demler, Phys. Rev. Lett. {\bf 96}, 190401 (2006).

\bibitem{Glassgold1963}
A. E. Glassgold, 
Phys. Rev. {\bf 132}, 2144 (1963).

\bibitem{Ueda2010} M. Ueda and Y. Kawaguchi, arXiv:1001.2072.

\bibitem{Herzberg1953}
G. Herzberg, {\em Molecular Spectra and Molecular Structure I: Spectra of Diatomic Molecules}, 2nd ed. (van Nostrand, New York, 1953).

\bibitem{Micheli2007}
A. Micheli, G. Pupillo, H. P. B\"uchler, and P. Zoller,
Phys. Rev. A {\bf 76}, 043604 (2007).

\bibitem{Jin-Ospelkaus}
D.~S.~Jin, S.~Ospelkaus, private communication.

\bibitem{Klawunn2010}
M. Klawunn, A. Pikovski, and L. Santos,
Phys. Rev. A {\bf 82}, 044701 (2010).

\bibitem{Shuler1968}
K.~E.~Shuler, 
in {\em Chemische Elementarprozesse}, edited by H.~Hartmann (Springer, Berlin, 1968).



\end{thebibliography}
\end{document}